\documentclass{Interspeech}

\interspeechcameraready 

\usepackage{amsmath,amssymb,amsfonts}
\usepackage{circuitikz}
\usepackage{subcaption}
\usepackage{tikz}
\usetikzlibrary{decorations.pathreplacing,calligraphy}
\usepackage{booktabs}
\usepackage{dcolumn}
\newcolumntype{d}[1]{D{.}{.}{#1}}
\usepackage{pgf}
\usepackage{pgfplots}
\usetikzlibrary{intersections}
\usetikzlibrary{positioning}
\usetikzlibrary{patterns}
\usepgfplotslibrary{fillbetween}
\usepackage{pifont}
\usepackage{adjustbox}
\usepackage[nolist, nohyperlinks]{acronym}
\usepackage{nicefrac}
\usepackage{hyperref}
\usepackage{cleveref}
\usepackage{stmaryrd}
\usepackage{setspace}
\usepackage{flushend}
\pgfplotsset{compat=1.18}

\acrodefplural{nns}[NNs]{neural networks}
\acrodefplural{irm}[IRMs]{ideal ratio masks}
\acrodefplural{ibm}[IBMs]{ideal binary masks}
\acrodefplural{doa}[DoAs]{directions of arrival}
\acrodefplural{rirs}[RIRs]{room impulse responses}

\begin{acronym}
\acro{stft}[STFT]{short-time Fourier transform}
\acro{rir}[RIR]{room impulse response}
\acro{istft}[iSTFT]{inverse short-time Fourier transform}
\acro{doa}[DoA]{direction of arrival}
\acro{irm}[IRM]{ideal ratio mask}
\acro{mvdr}[MVDR]{minimum-variance distortionless response}
\acro{nn}[NN]{neural network}
\acro{dnn}[DNN]{deep neural network}
\acro{gcc_phat}[GCC-PHAT]{generalized cross-correlation phase transform}
\acro{ssl}[SSL]{sound source localization}
\acro{srp_phat}[SRP-PHAT]{steered response power with phase transform}
\acro{mse}[MSE]{mean squared error}
\acro{mae}[MAE]{mean angular error}
\acro{sdr}[SDR]{signal-to-distortion ratio}
\acro{sisdr}[SI-SDR]{scale-invariant signal-to-distortion ratio}
\acro{pesq}[PESQ]{perceptual evaluation of speech quality}
\acro{wer}[WER]{word error rate}
\acro{pit}[PIT]{permutation-invariant training}
\acro{lbt}[LBT]{location-based training}
\acro{asr}[ASR]{automatic speech recognition}
\acro{bce}[BCE]{binary cross-entropy}
\acro{jnf}[JNF]{joint non-linear filter}
\acro{nn}[NN]{neural network}
\acro{dnn}[DNN]{deep neural network}
\acro{mccruse}[MC-CRUSE]{multi-channel convolutional recurrent U-net architecture for speech enhancement}
\acro{estoi}[ESTOI]{extended short-time objective intelligibility}
\acro{snr}[SNR]{signal-to-noise ratio}
\acro{iid}[i.i.d.]{independent and identically distributed}
\acro{wrt}[w.r.t.]{with respect to}
\acro{ssf}[SSF]{spatially selective filter}
\acro{cv}[CV]{constant velocity}
\acro{tse}[TSE]{target speaker extraction}
\acro{tst}[TST]{target speaker tracking}
\acro{tsl}[TSL]{target speaker localization}
\acro{dnsmos}[DNSMOS]{deep noise suppression mean opinion score}
\acro{map}[MAP]{maximum a posteriori}
\acro{kf}[KF]{Kalman filter}
\acro{pf}[PF]{particle filter}
\acro{das}[DaS]{delay-and-sum}
\acro{ae}[AE]{angular error}
\acro{acc}[ACC]{accuracy}
\end{acronym}

\definecolor{tab_blue}{RGB}{31, 119, 180}
\definecolor{tab_green}{RGB}{44, 160, 44}
\definecolor{tab_orange}{RGB}{255, 127, 14}
\definecolor{tab_purple}{RGB}{148, 103, 189}
\definecolor{tab_brown}{RGB}{165, 42, 38}
\definecolor{tab_pink}{RGB}{227, 119, 194}
\definecolor{tab_cyan}{RGB}{23, 190, 207}

\newcommand\tickWidth{0.5pt}
\newcommand\numRegions{\Theta}
\newcommand\onehot{z}
\newcommand\trackingWidth{4.5cm}
\newcommand\trackingHeight{3cm}
\newcommand\performanceWidth{4.5cm}
\newcommand\performanceHeight{3cm}
\newcommand\trackingTickWidth{0.5pt}
\newcommand\trackingMinorTick{1pt}
\newcommand\trackingMajorTick{2pt}
\newcommand\trackingMarkSize{1.5pt}
\newcommand\trackingFont{\footnotesize}

\title{Steering Deep Non-Linear Spatially Selective Filters for Weakly Guided Extraction of Moving Speakers in Dynamic Scenarios}

\author{Jakob}{Kienegger}
\author{Timo}{Gerkmann}

\affiliation{Signal Processing (SP)}{University of Hamburg}{Germany}
\email{jakob.kienegger@uni-hamburg.de, timo.gerkmann@uni-hamburg.de}

\usepackage{comment}

\definecolor{DarkRed}{RGB}{139,0,0}
\definecolor{CustomOrange}{RGB}{255,114,0}
\definecolor{DarkBlue}{RGB}{0,51,153}
\definecolor{LightBlue}{RGB}{102,153,255}
\definecolor{DarkGreen}{RGB}{0,120,0}

\begin{document}

\maketitle

\begin{abstract}
Recent speaker extraction methods using deep non-linear spatial filtering perform exceptionally well when the target direction is known and stationary.
However, spatially dynamic scenarios are considerably more challenging due to time-varying spatial features and arising ambiguities, e.g. when moving speakers cross.
While in a static scenario it may be easy for a user to point to the target's direction, manually tracking a moving speaker is impractical. 
Instead of relying on accurate time-dependent directional cues, which we refer to as strong guidance, in this paper we propose a weakly guided extraction method solely depending on the target’s initial position to cope with spatial dynamic scenarios.
By incorporating our own deep tracking algorithm and developing a joint training strategy on a synthetic dataset, 
we demonstrate the proficiency of our approach in resolving spatial ambiguities and even outperform a mismatched, but strongly guided extraction method.
\end{abstract}

\section{Introduction}
Given a real world recording containing a speech signal, the field of speech enhancement focuses on improving its quality and intelligibility by suppressing noise and reverberation artifacts.
Noise sources in form of additional interfering speech signals, such as in the so-called cocktail-party problem \cite{cherry53cocktail_party}, are especially challenging due to their non-stationary temporal-spectral properties.
Furthermore, such scenarios require additional information about the target speaker in order to resolve the ambiguity regarding which speaker to enhance and which speakers to suppress.
This particular problem setting in the realm of speech enhancement is referred to as \ac{tse} and defined by the task of identifying, disentangling and enhancing a target speech signal from a multi-speaker recording based on a speaker-specific cue. 
While a variety of cue modalities have been proposed for this purpose, see e.g. \cite{zmolikova23tse_overview} for an overview, spatial information becomes especially popular in the case of a stationary speaker and the availability of recordings from a microphone array.
Commonly under the assumption that contributing speech sources are spatially distinguishable by their azimuth orientation to the array, a \ac{ssf} can be employed and steered in the target's direction to extract the desired speech signal \cite{tesch24ssf_journal, briegleb23icospa, bohlender24sep_journal}.
Recently proposed data-driven implementations have gained a great deal of attention by demonstrating very high spatial selectivity with only a limited amount of microphones \cite{tesch24ssf_journal, briegleb23icospa, pandey12directional_speech_extraction, gu24rezero}.
By joint exploitation of spatial, temporal and spectral information, such deep non-linear \acp{ssf} drastically improve the enhancement performance over traditional linear filtering approaches as e.g. the \ac{mvdr} beamformer \cite{vary06mvdr}.
However, due to the resulting high spatial selectivity, they also result in an increased dependency on the provision of accurate 
directional information \cite{tesch24ssf_journal, pandey12directional_speech_extraction}.

A recording setup such as a seated conference meeting with a centered microphone array   \cite{chen20libricss} can legitimate the assumption of stationary and directionally distinct speaker locations.
The training procedure can even be adapted to increase the robustness of \acp{ssf} against small perturbations such as minor cue inaccuracies or head movement of a speaker \cite{tesch24ssf_journal}.
However, in more general settings such as e.g. the dinner party scenario considered in \cite{barker18chime5}, participants can move freely within the recording environment and the prior assumptions are no longer valid.
With the majority of data-driven \acp{ssf} utilizing temporal context to extract the target's speech signal \cite{tesch24ssf_journal, briegleb23icospa, bohlender24sep_journal, pandey12directional_speech_extraction}, the change from a stationary to a dynamic acoustic scenario bears a significant domain shift.
To counteract this effect, recent works include non-stationary speaker locations during training by abruptly changing their positions over the duration of the recording \cite{bohlender24sep_journal, he243stse}. 
Although spatialization with such piece-wise constant trajectories is far less computationally expensive than simulating continuous motion, their discontinuity leads to a poor generalization in real recordings as demonstrated in \cite{bohlender23tracking_continuous_movement} for the task of \ac{ssl}.
The arising necessity of time-dependent directional cues bears an additional challenge, since they are usually not available in practical applications. 
While \cite{tesch24ssf_journal, bohlender24sep_journal} have used an upstream \ac{ssl} to estimate the speaker locations for the task of speech separation, oracle information is necessary to disentangle speaker trajectories in case two speakers pass each other \cite{bohlender24sep_journal}.
Furthermore, during such indeterminate speaker arrangements a \ac{ssf} cannot differentiate the overlapping speech signals solely based on their relative azimuth orientations toward the microphone array, see the visualization in \cite{bohlender24sep_journal}.
Whereas enriching the cue with additional positional information \cite{pandey12directional_speech_extraction, gu24rezero} or changing its modality to include the target's temporal-spectral characteristics e.g. by utilizing an enrollment utterance \cite{he243stse, meng22lspex, li19mc_speakerbeam} can help to resolve the ambiguity, it greatly limits the applicability in real-world scenarios.

Instead of relying on accurate time-dependent directional cues, which we refer to as \textit{strong} guidance, we propose a \textit{weakly} guided \ac{tse} method which solely depends on the speaker's initial position.
By creating a synthetic dataset modeling continuous movement, we demonstrate the significant performance improvement of \acp{ssf} in dynamic scenarios and showcase their capability to resolve spatial ambiguities by learning to differentiate temporal-spectral patterns.
Finally, by introducing a \textit{joint} training strategy, we can not only report strong results combined with our own deep tracking \ac{nn}, but also paired with a very simple and inaccurate tracking algorithm.

\section{Problem definition}
\subsection{Target speaker extraction}\label{sec:tse}
Let the multi-channel observation $\mathbf{Y}$ be modeled as additive mixture of reverberant target speech $\mathbf{X}$ with noise $\mathbf{V}$.
The latter contains interfering speech sources as well as additive environmental and measurement noise.
In the \ac{stft} domain, this acoustic setup is resembled by
\begin{equation}\label{eq:signal_model}
    \mathbf{Y}_{tk} = \mathbf{X}_{tk} + \mathbf{V}_{tk} \, ,
\end{equation}
with $t$ and $k$ denoting frame and frequency bins respectively and vectorization conducted over the channel dimension.
In this work, we define the task of \ac{tse} as the recovery of the direct path propagated speech signal $S_{tk}$, also referred to as \textit{dry} speech signal \cite{tesch24ssf_journal}, at a predefined reference microphone. 
While a vast amount of methods have been proposed for this task, we will focus on deep \ac{nn} algorithms
computing a complex-valued mask $\mathcal{M}_{tk}$ to extract the target speech signal $\widehat{S}_{tk}$ by multiplication with a reference channel of the input $Y^0_{tk}$
\begin{equation}\label{eq:mask_extraction}
    \widehat{S}_{tk} = \mathcal{M}_{tk} Y^0_{tk} \, .
\end{equation}
The data-driven estimation of time-frequency mask $\mathcal{M}_{tk}$ conditioned on a target specific cue is the center of this work.

\subsection{Spatially selective filter}\label{sec:ssf}
Spatially selective filters (\acp{ssf}) exploit positional cues to distinguish the target's speech signal from interfering sources. 
Since accurate positional information in three-dimensional space is usually not available, only the relative azimuth orientation of the target speaker towards the microphone array is commonly utilized for this purpose \cite{tesch24ssf_journal, briegleb23icospa, bohlender24sep_journal}.
To some degree this can be legitimized in case the inter-microphone distance is negligible towards the distance between target speaker and recording array.
On the one hand the validity of the far-field approximation \cite{vary06mvdr} increases, reducing the influence of distance on the signal characteristics.
Furthermore, a change in speaker height only results in a slight change in elevation angle, which leaves the azimuth direction as the most distinct spatial feature.
However, there have been extensions proposed to incorporate information about elevation as well as distance into the extraction process \cite{pandey12directional_speech_extraction, gu24rezero}.
In this work we abide by common convention and constrain the cue on the azimuth angle, which we will refer to as \ac{doa} in the following.
Thus, the task of the \ac{ssf} lies in computing the time-frequency mask $\mathcal{M}_{tk}$ in \eqref{eq:mask_extraction} from the multi-channel input $\mathbf{Y}_{tk}$ conditioned on the \ac{doa} $\theta$. 
In most realizations, the latter is provided as a one-hot encoded vector $\mathbf{\onehot}$ corresponding to a specific spatial discretization in $\numRegions$ equidistant angular regions \cite{tesch24ssf_journal, bohlender24sep_journal, pandey12directional_speech_extraction}.
While in the majority of prior work this cue is time-independent and a stationary setting is assumed \cite{tesch24ssf_journal, briegleb23icospa, bohlender24sep_journal, pandey12directional_speech_extraction}, we explicitly include motion to target and interfering speakers, thus, necessitating time-dependent directional information $\theta_t$, as indicated by the index $t$.

\section{Proposed method}
\subsection{Weakly guided target speaker extraction}\label{sec:weak_guide}
\begin{figure}[t!]
\vspace*{-2.5pt}
\input{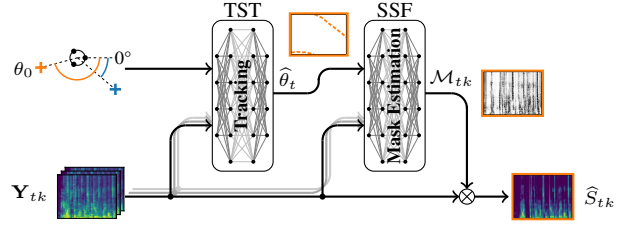}
\vspace*{-17.5pt}
\caption{Proposed weakly guided \acl{tse} (\acs{tse}) pipeline. In contrast to strongly guided \acs{tse}, our method
only requires the initial \ac{doa} $\theta_0$ to extract the speech signal $\widehat{S}_{tk}$.}
\vspace*{-10pt}
\label{fig:e2e_pipeline}
\end{figure}

Instead of relying on accurate \textit{continuous} prior directional information, we relax this constraint and only assume knowledge about the \textit{initial} starting direction $\theta_0$ of the target speaker.
In order to retain compatibility with conventional \acp{ssf} like \cite{tesch24ssf_journal, briegleb23icospa, bohlender24sep_journal, pandey12directional_speech_extraction}, we propose to incorporate an upstream tracking algorithm to estimate the evolution of the target's \ac{doa} $\theta_t$ based on $\theta_0$.
In analogy to \ac{tse}, we will refer to this task as \ac{tst} and avoid the term target speaker localization, as it is commonly paired with audio cues \cite{meng22lspex, li23gcc-speaker, chen24locselect}.
\Cref{fig:e2e_pipeline} displays the resulting \ac{tse} pipeline in form of a concatenation of an upstream \ac{tst} and downstream \ac{ssf} algorithm. 

\subsection{Target speaker tracking}\label{sec:target_speaker_tracking}
\begin{figure}[b!]
\centering
\input{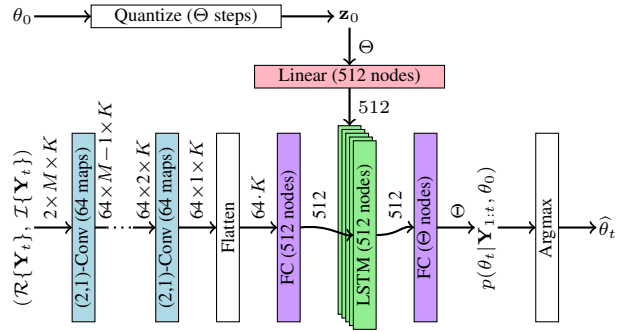}
\vspace*{-15pt}
\caption{Proposed deep \acl{tst} (\acs{tst}) conditioned on the initial \ac{doa} $\theta_0$. The network architecture is based on \cite{tesch24ssf_journal} and \cite{bohlender21ssl_temporal_context}. Colored layers indicate learnable parameters.}
\label{fig:ast_network}
\end{figure}
The general problem of estimating and tracking the direction of a moving source is a well-established field of research.
Classical model-based approaches such as \ac{kf} \cite{traa13wrapped_kalman_filter, murase05kalman_moving_speaker_tracking} or \ac{pf} \cite{dong20pf_doa_coprime, ward03basic_particle_filter} build on an underlying statistical framework and provide robust and computationally efficient algorithms. 
However, the performance of such algorithms substantially decreases in challenging noisy and reverberant scenarios \cite{zhang17distributed_particle_filter}. 
While improvements are possible by incorporation of data-driven methods \cite{revach22kalman_net, chen21diff_pf_cond_norm_flow}, in case enough training data is available, recent trends are moving toward fully neural approaches.
Especially deep recurrent architectures show great promise in the related field of \ac{ssl} by learning intricate dependencies between observation and the temporal evolution of the \ac{doa} \cite{bohlender21ssl_temporal_context, yang2022srp_dnn, yin22mimo_doanet}.
In this work we use the established causal \ac{ssl} model proposed by Bohlender et al. \cite{bohlender21ssl_temporal_context} as starting point and modify it for the purpose of \ac{tst}.
Due to a convolutional spatial preprocessing followed by a unidirectional LSTM layer, it is commonly referred to as CNN/LSTM \cite{bohlender24sep_journal, bohlender21ssl_temporal_context}. 
In its original implementation the \ac{doa} estimation is encoded in a classification problem in fixed angular steps, which is a common approach for data-driven \ac{ssl} algorithms \cite{tesch24ssf_journal, xiong15gcc_phat, papageorgiou21ssl}.
While the resulting spatial quantization can lead to estimation inaccuracies, in our specific use-case the output modality perfectly matches the one-hot encoding $\mathbf{\onehot}_t$ of the downstream \ac{ssf}.
To modify the architecture towards tracking a specific target speaker, we change the original output activation from Sigmoid to Softmax \cite{xiong15gcc_phat}, making it interpretable as posterior probability $p(\theta_t | \mathbf{Y}_{1:t}, \theta_0)$.
We then estimate the target's \ac{doa} $\widehat{\theta}_t$ according to the \ac{map} decision rule
\begin{equation}\label{eq:map}
    \widehat{\theta}_t = {\raisebox{3pt}{$\underset{\theta_t}{\mathrm{argmax}}$}} \ 
    p(\theta_t | \mathbf{Y}_{1:t}, \theta_0) \, .
\end{equation}
In order to inform the \ac{nn} about the initial \ac{doa} $\theta_0$, we choose to adopt the initialization technique employed in the FT-\acs{jnf} architecture \cite{tesch24ssf_journal}.
Specifically, we use the one-hot encoded \ac{doa} representation $\mathbf{\onehot}_t$ transformed by a linear layer directly as initialization of the LSTM in the \ac{tst} network.
Finally, we replace the original input features in \cite{bohlender21ssl_temporal_context} with the complex multi-channel spectrogram, matching both FT-\acs{jnf} and \ac{ssl} proposed in \cite{tesch24ssf_journal}.
The resulting deep \ac{tst} architecture is illustrated in \cref{fig:ast_network}.

\section{Experiments}
\subsection{Dataset}\label{sec:dataset}
\begin{figure}[t!]
\begin{tikzpicture}
    
\newcommand\specWidth{1.75cm}
\newcommand\specHeight{9mm}
\newcommand\specDist{1mm}
\newcommand\plotStartX{0}
\newcommand\plotStartY{0}
\newcommand\majorTick{0.75mm}
\newcommand\minorTick{0.5mm}
\newcommand\tickSize{7pt}
\newcommand\tickSkip{9pt}
\newcommand\timeDist{3.1mm} % 3.1
\newcommand\doaDist{1.75mm}
\newcommand\legendTick{5mm}
\newcommand\legendOff{2.7cm}
\newcommand\legendOffProp{2.7cm}
\newcommand\labelWidth{1pt}
\newcommand\descriptX{2.7mm}
\newcommand\descriptY{2.2mm}

% axis labels
\node[anchor=north] at (\plotStartX + 2 * \specWidth + 1* \specDist + - 0.5 * \specWidth+ 1 * \specDist, \plotStartY  - 0.5 * \specHeight - 2*\majorTick) {\footnotesize time [s]};
\node[anchor=center, rotate=90] at (\plotStartX - 0.86 * \specWidth, \plotStartY - 0.5 * \specHeight+ 1 * \specDist + 1*\specHeight) {\footnotesize DoA [°]};

% legend
% ground truth
\draw[color=tab_blue, line width=2 * \labelWidth, opacity=0.5] (\plotStartX - 0.5 * \specWidth, \plotStartY + 1.5 * \specHeight + 2.5* \specDist) --  (\plotStartX - 0.5 * \specWidth + \legendTick, \plotStartY + 1.5 * \specHeight + 2.5* \specDist);
\draw[color=tab_orange, line width=2 * \labelWidth, opacity=0.5] (\plotStartX - 0.5 * \specWidth, \plotStartY + 1.5 * \specHeight + 2.5* \specDist + \specDist) --  (\plotStartX - 0.5 * \specWidth + \legendTick, \plotStartY + 1.5 * \specHeight + 2.5* \specDist + \specDist);
\node[anchor=west] at (\plotStartX - 0.5 * \specWidth + \legendTick, \plotStartY + 1.5 * \specHeight + 3* \specDist) {\footnotesize Ground truth};

% particle filter
\draw[color=tab_blue, line width=\labelWidth, dash pattern=on 1pt off 1pt] (\plotStartX - 0.5 * \specWidth + \legendOff, \plotStartY + 1.5 * \specHeight + 2.5* \specDist) --  (\plotStartX - 0.5 * \specWidth + \legendTick + \legendOff, \plotStartY + 1.5 * \specHeight + 2.5* \specDist);
\draw[color=tab_orange, line width=\labelWidth, dash pattern=on 1pt off 1pt] (\plotStartX - 0.5 * \specWidth + \legendOff, \plotStartY + 1.5 * \specHeight + 3.5* \specDist) --  (\plotStartX - 0.5 * \specWidth + \legendTick + \legendOff, \plotStartY + 1.5 * \specHeight + 3.5* \specDist);
\node[anchor=west] at (\plotStartX - 0.5 * \specWidth + \legendTick + \legendOff, \plotStartY + 1.5 * \specHeight + 3* \specDist) {\footnotesize \acs{das}-\acs{pf} \cite{ward03basic_particle_filter}};

% proposed
\draw[color=tab_blue, line width=\labelWidth, dash pattern=on 3pt off 1pt] (\plotStartX - 0.5 * \specWidth + \legendOff + \legendOffProp, \plotStartY + 1.5 * \specHeight + 2.5* \specDist) --  (\plotStartX - 0.5 * \specWidth + \legendTick + \legendOff + \legendOffProp, \plotStartY + 1.5 * \specHeight + 2.5* \specDist);
\draw[color=tab_orange, line width=\labelWidth, dash pattern=on 3pt off 1pt] (\plotStartX - 0.5 * \specWidth + \legendOff + \legendOffProp, \plotStartY + 1.5 * \specHeight + 3.5* \specDist) --  (\plotStartX - 0.5 * \specWidth + \legendTick + \legendOff + \legendOffProp, \plotStartY + 1.5 * \specHeight + 3.5* \specDist);
\node[anchor=west] at (\plotStartX - 0.5 * \specWidth + \legendTick + \legendOff + \legendOffProp, \plotStartY + 1.5 * \specHeight + 3* \specDist) {\footnotesize Proposed};

% bottom row
\foreach \x in {0, ..., 3} {
    \pgfmathsetmacro{\filename}{\x}
    \node at (\plotStartX + \x * \specWidth + \x * \specDist, \plotStartY) {%
        \pgfimage[width=\specWidth, height=\specHeight]{images/doa_\filename.pdf} 
    };
    % time axis labels
    \foreach \y in {0, 2, 4} {
        \draw[line width=\tickWidth] (\plotStartX + \x * \specWidth + \x * \specDist + \y * \timeDist - 0.5 * \specWidth+ 1 * \specDist, \plotStartY - 0.5 * \specHeight+ 1* \specDist) --  (\plotStartX + \x * \specWidth + \x * \specDist + \y * \timeDist- 0.5 * \specWidth+ 1 * \specDist, \plotStartY - \majorTick - 0.5 * \specHeight+ 1 * \specDist);
        \pgfmathsetmacro{\time}{int(\y)} 
        \node[anchor=north] at (\plotStartX + \x * \specWidth + \x * \specDist + \y * \timeDist- 0.5 * \specWidth+ 1 * \specDist, \plotStartY  - 0.5 * \specHeight+ 1 * \specDist) {\fontsize{\tickSize}{\tickSkip}\selectfont \time};
    }
    \foreach \y in {1, 3, 5} {
        \draw[line width=\tickWidth] (\plotStartX + \x * \specWidth + \x * \specDist + \y * \timeDist - 0.5 * \specWidth+ 1 * \specDist, \plotStartY - 0.5 * \specHeight+ 1 * \specDist) --  (\plotStartX + \x * \specWidth + \x * \specDist + \y * \timeDist- 0.5 * \specWidth+ 1 * \specDist, \plotStartY - \minorTick - 0.5 * \specHeight+ 01 * \specDist);
    }
    % doa labels
    \foreach \y in {0, 2, 4} {
        \draw[line width=\tickWidth] (\plotStartX + \x * \specWidth + \x * \specDist - 0.5 * \specWidth+ 1 * \specDist - \majorTick, \plotStartY - 0.5 * \specHeight+ 1 * \specDist + \y * \doaDist) --  (\plotStartX + \x * \specWidth + \x * \specDist - 0.5 * \specWidth+ 1 * \specDist , \plotStartY - 0.5 * \specHeight+ 1 * \specDist+ \y * \doaDist);
        \ifthenelse{\x = 0}{ % this conditional throws an error, i dunno why
            \pgfmathsetmacro{\doa}{int(\y * 90 - 180)} 
            \node[anchor=east] at (\plotStartX + \x * \specWidth + \x * \specDist - 0.5 * \specWidth+ 1 * \specDist , \plotStartY - 0.5 * \specHeight+ 1 * \specDist+ \y * \doaDist) {\fontsize{\tickSize}{\tickSkip}\selectfont \doa};
        }{
        
        }
    }
    \foreach \y in {1, 3} {
        \draw[line width=\tickWidth] (\plotStartX + \x * \specWidth + \x * \specDist - 0.5 * \specWidth+ 1 * \specDist - \minorTick, \plotStartY - 0.5 * \specHeight+ 1 * \specDist + \y * \doaDist) --  (\plotStartX + \x * \specWidth + \x * \specDist - 0.5 * \specWidth+ 1 * \specDist , \plotStartY - 0.5 * \specHeight+ 1 * \specDist+ \y * \doaDist);
    }
}
% top row
\foreach \x in {0, ..., 3} {
    \pgfmathsetmacro{\filename}{int(\x + 4)}
    \node at (\plotStartX + \x * \specWidth + \x * \specDist, \plotStartY + \specHeight + \specDist) {%
        \pgfimage[width=\specWidth, height=\specHeight]{images/doa_\filename.pdf} 
    };
    
    % time axis labels
    \foreach \y in {0, 2, 4} {
        \draw[line width=\tickWidth] (\plotStartX + \x * \specWidth + \x * \specDist + \y * \timeDist - 0.5 * \specWidth+ 1 * \specDist, \plotStartY - 0.5 * \specHeight+ 1 * \specDist + \specHeight + \specDist) --  (\plotStartX + \x * \specWidth + \x * \specDist + \y * \timeDist- 0.5 * \specWidth+ 1 * \specDist, \plotStartY - \majorTick - 0.5 * \specHeight+ 1 * \specDist + \specHeight + \specDist);
    }
    \foreach \y in {1, 3, 5} {
        \draw[line width=\tickWidth] (\plotStartX + \x * \specWidth + \x * \specDist + \y * \timeDist - 0.5 * \specWidth+ 1 * \specDist, \plotStartY - 0.5 * \specHeight+ 1 * \specDist + \specHeight + \specDist) --  (\plotStartX + \x * \specWidth + \x * \specDist + \y * \timeDist- 0.5 * \specWidth+ 01 * \specDist, \plotStartY - \minorTick - 0.5 * \specHeight+ 1 * \specDist + \specHeight + \specDist);
    }
    % doa labels
    \foreach \y in {0, 2, 4} {
        \draw[line width=\tickWidth] (\plotStartX + \x * \specWidth + \x * \specDist - 0.5 * \specWidth+ 1 * \specDist - \majorTick, \plotStartY - 0.5 * \specHeight+ 1 * \specDist + \y * \doaDist+ \specHeight + \specDist) --  (\plotStartX + \x * \specWidth + \x * \specDist - 0.5 * \specWidth+ 1 * \specDist , \plotStartY - 0.5 * \specHeight+ 1 * \specDist+ \y * \doaDist+ \specHeight + \specDist);
        \ifthenelse{\x = 0}{ % this conditional throws an error, i dunno why
            \pgfmathsetmacro{\doa}{int(\y * 90 - 180)} 
            \node[anchor=east] at (\plotStartX + \x * \specWidth + \x * \specDist - 0.5 * \specWidth+ 1 * \specDist , \plotStartY - 0.5 * \specHeight+ 1 * \specDist+ \y * \doaDist+ \specHeight + \specDist) {\fontsize{\tickSize}{\tickSkip}\selectfont \doa};
        }{
        
        }
    }
    \foreach \y in {1, 3} {
        \draw[line width=\tickWidth] (\plotStartX + \x * \specWidth + \x * \specDist - 0.5 * \specWidth+ 1 * \specDist - \minorTick, \plotStartY - 0.5 * \specHeight+ 1 * \specDist + \y * \doaDist+ \specHeight + \specDist) --  (\plotStartX + \x * \specWidth + \x * \specDist - 0.5 * \specWidth+ 1 * \specDist , \plotStartY - 0.5 * \specHeight+ 1 * \specDist+ \y * \doaDist+ \specHeight + \specDist);
    }
}

% descriptions
% \node at (\plotStartX + 0 * \specWidth + 0 * \specDist - 0.5 * \specWidth + \descriptX, \plotStartY + 0.5*\specHeight + \specDist + \descriptY) {%
%     \scriptsize (a)
% };
% \node at (\plotStartX + 1 * \specWidth + 1 * \specDist - 0.5 * \specWidth + \descriptX, \plotStartY + 0.5*\specHeight + \specDist + \descriptY) {%
%     \scriptsize (b)
% };
% \node at (\plotStartX + 2 * \specWidth + 2 * \specDist - 0.5 * \specWidth + \descriptX, \plotStartY + 0.5*\specHeight + \specDist + \descriptY) {%
%     \scriptsize (c)
% };
% \node at (\plotStartX + 3 * \specWidth + 3 * \specDist - 0.5 * \specWidth + \descriptX, \plotStartY + 0.5*\specHeight + \specDist + \descriptY) {%
%     \scriptsize (d)
% };
% \node at (\plotStartX + 0 * \specWidth + 0 * \specDist - 0.5 * \specWidth + \descriptX, \plotStartY + 0.5*\specHeight  - \descriptY) {%
%     \scriptsize (e)
% };
% \node at (\plotStartX + 1 * \specWidth + 1 * \specDist - 0.5 * \specWidth + \descriptX, \plotStartY + 0.5*\specHeight  - \descriptY) {%
%     \scriptsize (f)
% };
% \node at (\plotStartX + 2 * \specWidth + 2 * \specDist - 0.5 * \specWidth + \descriptX, \plotStartY + 0.5*\specHeight  - \descriptY) {%
%     \scriptsize (g)
% };
% \node at (\plotStartX + 3 * \specWidth + 3 * \specDist - 0.5 * \specWidth + \descriptX, \plotStartY + 0.5*\specHeight  - \descriptY) {%
%     \scriptsize (h)
% };
\end{tikzpicture}
\vspace*{-17.5pt}
\caption{
\ac{doa} estimation with \acl{tst} (\acs{tst}) methods presented in \cref{sec:nn_training}. Selected trajectories correspond to \eqref{eq:motion_model} with an expected displacement $\mathbb{E}\hspace*{-2pt}\left\{ |\Delta \theta_t| \right\}$ of {\small $\frac{180^\circ}{5\mathrm{s}}$}.
}
\label{fig:moving_examples}
\end{figure}
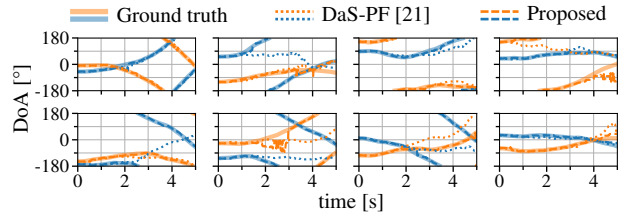

For training and evaluation we create a synthetic dataset with reverberant two-speaker mixtures.
Specifically, we use utterances from the LibriSpeech \cite{panayotov15librispeech} corpus, convolve them with simulated \acp{rirs} based on the image method \cite{allen79image_method} and then combine them according to the recipe in Libri2Mix \cite{cosentino20librimix}.
Since no additional noise is present in the mixture, the interfering signal $\mathbf{V}_{tk}$ in \eqref{eq:signal_model} consists of the early and late reflections of the target speech signal and the reverberant interfering speech.
The acoustic room setup closely follows the work from Tesch et al. \cite{tesch24ssf_journal}. 
Hence, we choose a circular array of $10\,\mathrm{cm}$ diameter as recording device, which consists of three uniformly spaced omnidirectional microphones.
We adhere to the reverberation times between 0.2\,s and 0.5\,s as well as the same simulation setup for the shoe-box rooms.
At the start of each mixture, target and interfering speaker are separated by at least $10^\circ$ and at a distance between $0.8$ to $1.2\,\mathrm{m}$ from the array.
The distance from the array as well as the height of the speaker is kept constant over the duration of each mixture while the azimuth direction may vary.
For this purpose we employ a \acl{cv} motion model \cite{rong03survey_target_tracking} defined as
\begin{align}
    \begin{bmatrix}
    \theta_{t+1} \\
    \dot{\theta}_{t+1}
    \end{bmatrix} &= \begin{bmatrix}
        1 & \hspace{-2mm}\Delta T \\
        0 & 1
    \end{bmatrix} \begin{bmatrix}
    \theta_t \\
    \dot{\theta}_t
    \end{bmatrix} + \begin{bmatrix}
        \Delta T^2/ 2 \\
        \Delta T
    \end{bmatrix} \nu_t \, , \label{eq:motion_model}
\intertext{with the perturbation $\nu_t$ being a white, zero mean Gaussian process of variance $\sigma^2$ and implicitly taking the wrapping of $\theta_t$ into account.
Since there is no constraint on the azimuth movement, intersecting trajectories of target and interfering speakers are allowed and depending on the perturbation $\nu_t$, more or less likely to happen.
To obtain a physically meaningful parametrization, we derive the distribution of the angular displacement $\Delta \theta_t$ between start and current frame $t$.
Since this quantity is not affected by wrapping effects, sequentially evaluating the above linear state-space model leads to the Gaussian distribution}
    \Delta \theta_t &\sim \mathcal{N}\left( 0, \frac{{\Delta T}^4}{12} \left(4t^3 - t\right) \sigma^2\right) \, . \label{eq:delta_theta}
\intertext{
Due to its symmetry, the expected absolute displacement $\mathbb{E}\hspace*{-2pt}\left\{ |\Delta \theta_t| \right\}$ can be expressed through the closed-form solution}
    \mathbb{E}\hspace*{-2pt}\left\{ |\Delta \theta_t| \right\} &= \frac{\Delta T^2}{\sqrt{6 \pi}} \sqrt{4t^3 - t} \, \sigma  \, .\label{eq:expected_displacement}
\end{align}
In the following, we will determine the standard deviation $\sigma$ of the motion model's perturbation $\nu_t$ in \eqref{eq:motion_model} from a given mean absolute deviation $\mathbb{E}\hspace*{-2pt}\left\{ |\Delta \theta_t| \right\}$ after $t$ frames using \eqref{eq:expected_displacement}.
The trajectory time step $\Delta T$ of the \acs{rir} simulations is aligned with the \ac{stft} parametrization, for which we use a square-root Hann window \cite{shimauchi14hann_window} of length $32\,\mathrm{ms}$ and $16\,\mathrm{ms}$ hop-size. \Cref{fig:moving_examples} illustrates the resulting motion dynamics corresponding to an expected angular displacement $\mathbb{E}\hspace*{-2pt}\left\{ |\Delta \theta_t| \right\}$ of {\small $\frac{180^\circ}{5\mathrm{s}}$}.

\subsection{Network architectures and training details}\label{sec:nn_training}

\hspace*{\parindent}\textit{\textbf{Spatially selective filter}} 
To test our proposed methods with a strong \ac{ssf} baseline we choose FT-\acs{jnf} proposed by Tesch et al. in \cite{tesch24ssf_journal}.
Due to its exceptional spatial filtering capabilities paired with a conceptually simple and lightweight architecture, it has seen great popularity in a variety of applications \cite{wechsler24directional_filtering_directivity_control, briegleb24constrained_vs_unconstained_filtering, lentz24ftjnf_head_rotation}.
In order to retain causality, we slightly deviate from \cite{tesch24ssf_journal} and replace the last B-LSTM layer by an unidirectional LSTM of twice its initial parameters to roughly preserve the intended model size.
We train FT-\acs{jnf} with an initial learning rate of $10^{-3}$ and exponential decay of $10^{-2}$ according to the joint time and frequency domain loss function in \cite{tesch24ssf_journal}.

\textit{\textbf{Target speaker tracking}}
To cope with the challenging motion dynamics in our dataset, we utilize our proposed architecture in \cref{sec:target_speaker_tracking} as upstream \ac{tst} method with the parametrization shown in \cref{fig:ast_network}.
We train the \ac{nn} to minimize the cross-entropy towards the one-hot encoded ground truth \ac{doa} $\mathbf{\onehot}_t$ with an initial learning rate of $10^{-4}$ and exponential decay of $10^{-2}$.
To give a comparison with a traditional approach, we choose a \ac{pf} due to its popularity and algorithmic simplicity.
Specifically, we employ the algorithm in \cite{ward03basic_particle_filter} using the output power of a \ac{das} beamformer as likelihood, which we will refer to as \acs{das}-\acs{pf} in the following. 
To strengthen this baseline, we employ the motion model in \eqref{eq:motion_model} and match the perturbation's standard deviation $\sigma$ to the dataset.

\textit{\textbf{Training}}
All \acp{nn} are trained on mixtures of $5\,\mathrm{s}$ length while resampling the randomized acoustic setup from \cref{sec:dataset} including the speaker trajectories in each step.
The latter becomes feasible due to the GPU accelerated implementation of the image method in \cite{diaz18gpu_rir}.
We use spatial discretization of $2^\circ$ which results in a total number of 180 spatial regions $\numRegions$.
We employ two different training strategies for our proposed \ac{tse} pipeline in \cref{fig:e2e_pipeline}.
In the first we train the \acp{nn} independently for a total of 100 and 50 epochs regarding the \ac{ssf} and \ac{tst} architectures respectively.
The second strategy is to train the pipeline in an end-to-end fashion, consequently, we drop the argmax operation in \eqref{eq:map} to retain differentiability.
For comparability, we pre-train the \ac{ssf} and \ac{tst} \acp{nn} independently for 75 and 25 epochs and then continue for the remaining 25 epochs on their concatenation regarding the loss of the FT-\acs{jnf}.
 
\section{Results}
\begin{table}[t!]
 \caption{
 Performance of \acl{tse} (\acs{tse}) pipelines with FT-\acs{jnf} \cite{tesch24ssf_journal} and varying tracking and training strategies for expected \acs{doa} shifts $\mathbb{E}\hspace*{-2pt}\left\{ |\Delta \theta_t| \right\}$ of {\small $\frac{0^\circ}{5\mathrm{s}}$}$/${\small $\frac{180^\circ}{5\mathrm{s}}$}$/${\small $\frac{360^\circ}{5\mathrm{s}}$}.
}
\label{tab:results}
\resizebox{\columnwidth}{!}{
\setlength{\tabcolsep}{2pt}
\renewcommand{\arraystretch}{0.9}
\footnotesize
\begin{tabular}{cllccc}
 \toprule[1.5pt]
 & \multicolumn{2}{c}{\textbf{Pipeline}} & \multicolumn{3}{c}{\textbf{Performance}\,\big[{\small $\frac{0^\circ}{5\mathrm{s}}$}$/${\small $\frac{180^\circ}{5\mathrm{s}}$}$/${\small $\frac{360^\circ}{5\mathrm{s}}$}\big]} \\[1pt] \cmidrule(lr){2-3} \cmidrule(lr){4-6} 
\textbf{ID} & Tracking & Training & \acs{sisdr}\,[dB]\,$\uparrow$ & \acs{pesq}\,[-0.5$-$4.5]\,$\uparrow$ & \acs{estoi}\,[\%]\,$\uparrow$  \\ \midrule
(0) &  $-$ & $-$ &  -6.81$/$-6.82$/$-6.83 & 1.13$/$1.13$/$1.12 & 43.9$/$43.2$/$42.9 \\ [-1pt] \cmidrule(lr){1-6}
(1) & Oracle & Static  &\textbf{5.41}$/$2.18$/$1.72 & \textbf{1.99}$/$1.61$/$1.56 & \textbf{75.8}$/$69.3$/$67.9 \\
(2) & Oracle & Dyn.(ours) & 4.71$/$\textbf{4.27}$/$\textbf{4.02} & 1.93$/$\textbf{1.84}$/$\textbf{1.80} & 74.3$/$\textbf{72.6}$/$\textbf{71.8}\\ [-1pt] \cmidrule(lr){1-6}
(3) & \acs{das}-\acs{pf}\cite{ward03basic_particle_filter} & Static  & \textbf{5.41}$/$-1.84$/$-4.38 & \textbf{1.99}$/$1.33$/$1.20 & \textbf{75.8}$/$52.5$/$40.1 \\ 
(4) & \acs{das}-\acs{pf}\cite{ward03basic_particle_filter} & Dyn.(ours) & 4.71$/$-1.69$/$-4.20 & 1.93$/$1.28$/$1.21 & 74.3$/$49.0$/$44.1 \\
(5) & \acs{das}-\acs{pf}\cite{ward03basic_particle_filter} & Joint(ours) & 4.27$/$\textbf{1.97}$/$\textbf{0.39} & 1.86$/$\textbf{1.54}$/$\textbf{1.42}& 72.6$/$\textbf{64.1}$/$\textbf{59.1} \\ [-1pt] \cmidrule(lr){1-6}
(6) & Proposed & Static  & \textbf{5.21}$/$0.73$/$-0.21 & \textbf{1.98}$/$1.56$/$1.49 & \textbf{75.5}$/$67.5$/$64.5  \\ 
(7) & Proposed & Dyn.(ours) & 4.41$/$2.37$/$1.38 & 1.93$/$1.75$/$1.65 & 74.0$/$70.2$/$\textbf{67.2}\\ 
(8) & Proposed & Joint(ours) &  4.56$/$\textbf{3.44}$/$\textbf{2.21} & 1.94$/$\textbf{1.77}$/$\textbf{1.66}& 74.0$/$\textbf{70.3}$/$66.5 \\ 
 \bottomrule[1.5pt]
\end{tabular}
}
\end{table}
\begin{figure}[b!]
\begin{minipage}{0.48\columnwidth}
    \centering
    \begin{tikzpicture}
    \begin{axis}[
         every axis/.append style={
            font=\trackingFont, 
        },
            ylabel={$\Delta$\acs{sisdr}\,[dB]\,$\rightarrow$},
            xlabel={exp. \acs{doa} shift $\mathbb{E}\hspace*{-2pt}\left\{ |\Delta \theta_t| \right\}$ \big[{\scriptsize $\frac{^\circ}{5\mathrm{s}}$}\big]},
            width=\performanceWidth,
            height=\performanceHeight,
            xmin=0, xmax=360,
            ymin=6, ymax=16,
            xtick={0, 90, 180, 270, 360},
            xticklabels={0, 90, 180, 270, 360},
            ytick={6, 8, 10, 12, 14, 16},
            yticklabels={6, 8, 10, 12, 14, 16},
            xticklabel style={yshift=2pt}, 
            yticklabel style={xshift=2pt},
            ylabel style={yshift=1mm, xshift=-1mm, rotate=0, anchor=center},
            xlabel style={yshift=-1.5mm, rotate=0, anchor=center},
            major tick length=\trackingMajorTick,
            minor tick length=\trackingMinorTick,
            minor y tick num=1,
            minor x tick num=2,
            ytick pos=left,
            xtick pos=bottom,
            tick align=outside,
            tick style={line width=\trackingTickWidth, color=black},
            grid=both,
            legend style={
                draw=none,
                at={(0.45, 1.05)},
                legend columns=1,
                cells={align=left, anchor=west},
                anchor=south,
                font=\trackingFont,
                % draw=black, % Draw a box around the legend
                fill=none, % Optional: Fill the box with white color
            },
        ]

        % Oracle static train
        \addplot[
            color=tab_purple,
            mark=square*,
            mark size=\trackingMarkSize,
            solid,
        ]
        coordinates {
            (0, 12.423932075500488)
            (30, 10.56002426147461)
            (60, 9.939882278442383)
            (90, 9.574124336242676)
            (120, 9.313301086425781)
            (150, 9.116613388061523)
            (180, 8.999454498291016)
            (210, 8.896795272827148)
            (240, 8.791468620300293)
            (270, 8.704497337341309)
            (300, 8.634811401367188)
            (330, 8.591821670532227)
            (360, 8.553266525268555)
        };
        \addlegendentry{Oracle/Static\,(1)};

        % Oracle dynamic train
        \addplot[
            color=tab_green, mark=triangle*,
            mark size=\trackingMarkSize,
            solid,
        ]
        coordinates {
            (0, 11.521096229553223)
            (30, 11.505714416503906)
            (60, 11.38517951965332)
            (90, 11.290679931640625)
            (120, 11.217970848083496)
            (150, 11.14218521118164)
            (180, 11.086888313293457)
            (210, 11.054316520690918)
            (240, 10.995222091674805)
            (270, 10.959881782531738)
            (300, 10.915522575378418)
            (330, 10.879749298095703)
            (360, 10.849127769470215)
        };
        \addlegendentry{Oracle/Dyn.(ours)\,(2)};

         % oracle static error
        \addplot [name path=st_lower, fill=none, draw=none]
        coordinates{
            (0, 9.6580069065094)
            (30, 7.637639284133911)
            (60, 7.149125576019287)
            (90, 6.900210857391357)
            (120, 6.741170644760132)
            (150, 6.6131908893585205)
            (180, 6.5206029415130615)
            (210, 6.504053831100464)
            (240, 6.448350191116333)
            (270, 6.385734558105469)
            (300, 6.33931040763855)
            (330, 6.334101676940918)
            (360, 6.325042247772217)
        };
        \addplot [name path=st_upper, fill=none, draw=none]
        coordinates{
            (0, 15.189857244491577)
            (30, 13.482409238815308)
            (60, 12.730638980865479)
            (90, 12.248037815093994)
            (120, 11.88543152809143)
            (150, 11.620035886764526)
            (180, 11.47830605506897)
            (210, 11.289536714553833)
            (240, 11.134587049484253)
            (270, 11.023260116577148)
            (300, 10.930312395095825)
            (330, 10.849541664123535)
            (360, 10.781490802764893)
        };
        \addplot[color=tab_purple, fill opacity=0.2] fill between[of=st_lower and st_upper];

         % oracle dynamic error
        \addplot [name path=st_lower, fill=none, draw=none]
        coordinates{
            (0, 8.601692914962769)
            (30, 8.710784196853638)
            (60, 8.590544939041138)
            (90, 8.525157451629639)
            (120, 8.473414182662964)
            (150, 8.394545555114746)
            (180, 8.302605628967285)
            (210, 8.300345420837402)
            (240, 8.255997896194458)
            (270, 8.221242666244507)
            (300, 8.175750017166138)
            (330, 8.162709951400757)
            (360, 8.149237394332886)
        };
        \addplot [name path=st_upper, fill=none, draw=none]
        coordinates{
            (0, 14.440499544143677)
            (30, 14.300644636154175)
            (60, 14.179814100265503)
            (90, 14.056202411651611)
            (120, 13.962527513504028)
            (150, 13.889824867248535)
            (180, 13.871170997619629)
            (210, 13.808287620544434)
            (240, 13.734446287155151)
            (270, 13.69852089881897)
            (300, 13.655295133590698)
            (330, 13.59678864479065)
            (360, 13.549018144607544)
        };
        \addplot[color=tab_green, fill opacity=0.2] fill between[of=st_lower and st_upper];

    \end{axis}

\end{tikzpicture}
\end{minipage}\hfill
\begin{minipage}{0.48\columnwidth}
    \centering
    \begin{tikzpicture}
    \begin{axis}[
         every axis/.append style={
            font=\trackingFont, 
        },
            ylabel={$\Delta$\acs{sisdr}\,[dB]\,$\rightarrow$},
            xlabel={exp. \acs{doa} shift $\mathbb{E}\hspace*{-2pt}\left\{ |\Delta \theta_t| \right\}$ \big[{\scriptsize $\frac{^\circ}{5\mathrm{s}}$}\big]},
            width=\performanceWidth,
            height=\performanceHeight,
            xmin=0, xmax=360,
            ymin=-4, ymax=16,
            xtick={0, 90, 180, 270, 360},
            xticklabels={0, 90, 180, 270, 360},
            ytick={-4, 0, 4, 8, 12, 16},
            yticklabels={-4, 0, 4, 8, 12, 16},
            xticklabel style={yshift=2pt}, 
            yticklabel style={xshift=2pt},
            ylabel style={yshift=1mm, xshift=-1mm, rotate=0, anchor=center},
            xlabel style={yshift=-1.5mm, rotate=0, anchor=center},
            major tick length=\trackingMajorTick,
            minor tick length=\trackingMinorTick,
            minor y tick num=1,
            minor x tick num=2,
            ytick pos=left,
            xtick pos=bottom,
            tick align=outside,
            tick style={line width=\trackingTickWidth, color=black},
            grid=both,
            legend style={
                draw=none,
                at={(0.45, 1.05)},
                legend columns=1,
                cells={align=left, anchor=west},
                anchor=south,
                font=\trackingFont,
                % draw=black, % Draw a box around the legend
                fill=none, % Optional: Fill the box with white color
            },
        ]

        % pf dynamic
        \addplot[
            color=tab_green,
            mark=triangle*,
            mark size=\trackingMarkSize,
            dashed,
        ]
        coordinates {
            (0, 11.52)
            (30, 9.26)
            (60, 7.84)
            (90, 6.77)
            (120, 6.24)
            (150, 5.53)
            (180, 5.13)
            (210, 4.63)
            (240, 4.13)
            (270, 3.73)
            (300, 3.38)
            (330, 3.09)
            (360, 2.63)
        };
        \addlegendentry{\acs{das}-\acs{pf}\cite{ward03basic_particle_filter}/Dyn.(ours)\,(4)};

        % pf joint
        \addplot[
            color=tab_brown, mark=pentagon*,
            mark size=\trackingMarkSize,
            solid,
        ]
        coordinates {
            (0, 11.08)
            (30, 10.88)
            (60, 10.19)
            (90, 9.77)
            (120, 9.42)
            (150, 9.09)
            (180, 8.79)
            (210, 8.50)
            (240, 8.26)
            (270, 7.99)
            (300, 7.71)
            (330, 7.43)
            (360, 7.20)
        };
        \addlegendentry{\acs{das}-\acs{pf}\cite{ward03basic_particle_filter}/Joint(ours)\,(5)};

         % pf dynamic error
        \addplot [name path=st_lower, fill=none, draw=none]
        coordinates{
            (0, 8.60)
            (30, 5.14)
            (60, 3.01)
            (90, 1.62)
            (120, 0.88)
            (150, 0.03)
            (180, -0.78)
            (210, -1.30)
            (240, -1.70)
            (270, -2.30)
            (300, -2.71)
            (330, -2.85)
            (360, -3.09)
        };
        \addplot [name path=st_upper, fill=none, draw=none]
        coordinates{
            (0, 14.440499544143677)
            (30, 13.39)
            (60, 12.57)
            (90, 11.97)
            (120, 11.43)
            (150, 11.11)
            (180, 10.70)
            (210, 10.34)
            (240, 9.97)
            (270, 9.74)
            (300, 9.35)
            (330, 9.01)
            (360, 8.80)
        };
        \addplot[color=tab_green, fill opacity=0.2] fill between[of=st_lower and st_upper];

         % pf joint error
        \addplot [name path=st_lower, fill=none, draw=none]
        coordinates{
            (0, 8.08)
            (30, 7.43)
            (60, 6.76)
            (90, 6.11)
            (120, 5.56)
            (150, 5.0)
            (180, 4.57)
            (210, 4.02)
            (240, 3.65)
            (270, 3.24)
            (300, 2.80)
            (330, 2.37)
            (360, 2.03)
        };
        \addplot [name path=st_upper, fill=none, draw=none]
        coordinates{
            (0, 14.08)
            (30, 13.91)
            (60, 13.62)
            (90, 13.47)
            (120, 13.31)
            (150, 13.17)
            (180, 13.07)
            (210, 12.99)
            (240, 12.89)
            (270, 12.68)
            (300, 12.62)
            (330, 12.52)
            (360, 12.39)
        };
        \addplot[color=tab_brown, fill opacity=0.2] fill between[of=st_lower and st_upper];

    \end{axis}

\end{tikzpicture}
\end{minipage}
\vspace*{-17.5pt}
\caption{Selected \acs{tse} configurations evaluated with a metric sensitive to distortions. Shaded areas indicate std. deviation.}
\label{fig:sisdr_performance}
\end{figure}
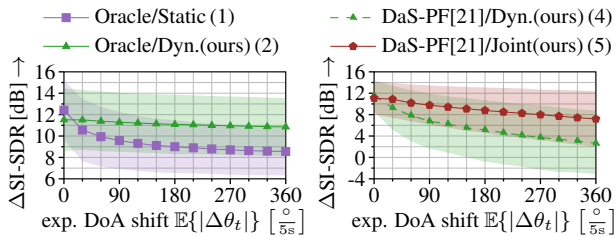
We evaluate the \ac{tse} performance w.r.t. distortions, perceptual quality and intelligibility via the metrics \acs{sisdr} \cite{roux19sisdr_half_baked}, \acs{pesq} \cite{rix01pesq} and \acs{estoi} \cite{jensen16estoi} respectively.
\Cref{tab:results} displays the results using FT-\acs{jnf} with different tracking and training strategies for three levels of speaker dynamics, complemented by
\cref{fig:sisdr_performance} regarding \acs{sisdr} improvement with a refined resolution.

\begin{figure}[h!]
\begin{tikzpicture}
    
\newcommand\specWidth{3.25cm}
\newcommand\specHeight{11mm}
\newcommand\cbarHeight{5mm}
\newcommand\cbarDist{6.67mm}
\newcommand\specDist{0mm} % 2
\newcommand\specDistY{-0.5mm}
\newcommand\plotStartX{0mm}
\newcommand\plotStartY{0mm}
\newcommand\majorTick{0.75mm}
\newcommand\minorTick{0.5mm}
\newcommand\timeDist{2.525mm}
\newcommand\doaDist{2.12mm}
\newcommand\startX{1.1mm}
\newcommand\startY{-4.2mm}
\newcommand\tickSize{7pt}
\newcommand\tickSkip{9pt}
\newcommand\legendOffX{15.6mm}
\newcommand\legendOffY{2.75mm}
\newcommand\legendOffDoubleY{1.5mm}
\newcommand\legendFont{\scriptsize}

% left row
% dynamic oracle
\node at (\plotStartX ,\plotStartY) {%
    \pgfimage[height=\specHeight, width=\specWidth]{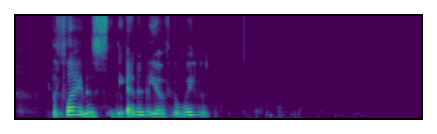} 
};
\node[color=white, anchor=east, align=right] at (\plotStartX + \legendOffX ,\plotStartY + \legendOffDoubleY) {%
    \legendFont Oracle/Dyn. \\[-4pt] \legendFont (ours)\,(2)
};
% static oracle
\node at (\plotStartX ,\plotStartY + \specHeight + \specDistY) {%
    \pgfimage[height=\specHeight, width=\specWidth]{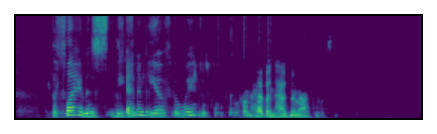} 
};
\node[color=white, anchor=east] at (\plotStartX + \legendOffX ,\plotStartY + \legendOffY+ \specHeight + \specDistY) {%
    \legendFont Oracle/Static\,(1)
};
% clean
\node at (\plotStartX ,\plotStartY + 2* \specHeight + 2*\specDistY) {%
    \pgfimage[height=\specHeight, width=\specWidth]{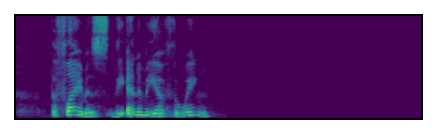} 
};
\node[color=white, anchor=east] at (\plotStartX + \legendOffX ,\plotStartY + \legendOffY+ 2* \specHeight + 2*\specDistY) {%
    \legendFont Ground\,truth
};
% doa
\node at (\plotStartX ,\plotStartY + 3* \specHeight + 3*\specDistY) {%
    \pgfimage[height=\specHeight, width=\specWidth]{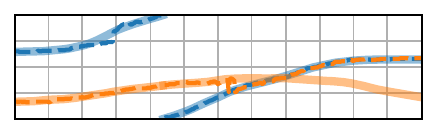} 
};
% right row
% joint
\node at (\plotStartX + \specWidth + \specDist ,\plotStartY) {%
    \pgfimage[height=\specHeight, width=\specWidth]{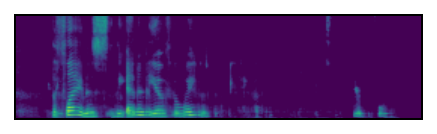} 
};
\node[color=white, anchor=east, align=right] at (\plotStartX + \legendOffX + \specWidth + \specDist,\plotStartY +\legendOffDoubleY) {%
    \legendFont Proposed/Joint \\[-4pt] \legendFont (ours)\,(8)
};
% dynamic concat
\node at (\plotStartX  + \specWidth + \specDist,\plotStartY + \specHeight + \specDistY) {%
    \pgfimage[height=\specHeight, width=\specWidth]{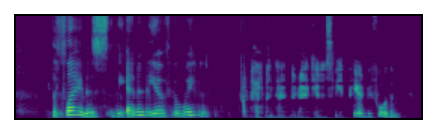} 
};
\node[color=white, anchor=east, align=right] at (\plotStartX + \legendOffX   + \specWidth + \specDist,\plotStartY + \specHeight + \specDistY + \legendOffDoubleY) {%
    \legendFont Proposed/Dyn.\\[-4pt] \legendFont (ours)\,(7)
};
% static concat
\node at (\plotStartX  + \specWidth + \specDist,\plotStartY + 2* \specHeight + 2*\specDistY) {%
    \pgfimage[height=\specHeight, width=\specWidth]{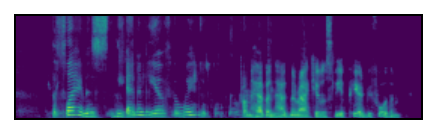} 
};
\node[color=white, anchor=east, align=right] at (\plotStartX + \legendOffX + \specWidth + \specDist,\plotStartY + 2* \specHeight + 2*\specDistY + \legendOffDoubleY) {%
    \legendFont Proposed/\\[-4pt] \legendFont Static\,(6)
};
% noisy
\node at (\plotStartX  + \specWidth + \specDist,\plotStartY + 3* \specHeight + 3*\specDistY) {%
    \pgfimage[height=\specHeight, width=\specWidth]{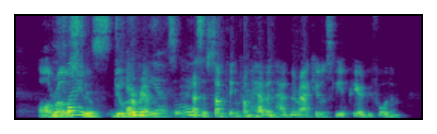} 
};
\node[color=white, anchor=east] at (\plotStartX + \legendOffX + \specWidth + \specDist,\plotStartY + 3* \specHeight + 3*\specDistY + \legendOffY) {%
    \legendFont Unprocessed\,(0)
};

% colormap
\pgfmathsetmacro{\cbarWidth}{4*\specHeight + 3* \specDistY - 2}
\node[rotate=90, anchor=center] at (\plotStartX + 1*\specWidth + 1.65*\specHeight, 0.5*\cbarWidth pt - 0.5*\specHeight + 1) {%
    \pgfimage[height=\cbarHeight, width=\cbarWidth pt]{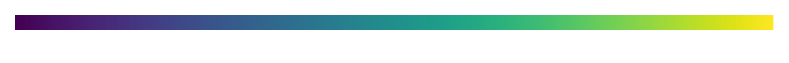} 
};
\foreach \c in {0, ..., 6}{
    \draw[line width=\tickWidth] (\plotStartX + 1*\specWidth + 1.65*\specHeight + 1*\specDist -0.1mm,\plotStartY + \startY+  \c * \cbarDist) --  (\plotStartX + 1*\specWidth + 1.65*\specHeight + 1*\specDist + \majorTick, \plotStartY + \startY+ \c * \cbarDist);
    \pgfmathsetmacro{\clabel}{int(\c * 10 - 60)}
    \node[anchor=west] at (\plotStartX + 1*\specWidth + 1.65*\specHeight + 1*\specDist, \plotStartY + \startY+ \c * \cbarDist){\fontsize{\tickSize}{\tickSkip}\selectfont \clabel};
    \ifthenelse{\c < 6}{
        \draw[line width=\tickWidth] (\plotStartX + 1*\specWidth + 1.65*\specHeight + 1*\specDist -0.1mm,\plotStartY + \startY+  \c * \cbarDist + 0.5*\cbarDist) --  (\plotStartX + 1*\specWidth + 1.65*\specHeight + 1*\specDist + \minorTick, \plotStartY + \startY+ \c * \cbarDist + 0.5*\cbarDist);
    }{}
}
\node[anchor=north] at (\plotStartX + 1*\specWidth + 1.65*\specHeight+4,\plotStartY + \startY - 2) {\footnotesize [dB]};

% axis labels
\node[anchor=north, rotate=90] at (\plotStartX + \startX - 0.5 * \specWidth - 0.85*\specHeight, 3 * \specDistY + \plotStartY + \startY +3 * \specHeight + 2 * \doaDist){\footnotesize DoA [°]};
\node[anchor=south, rotate=90] at (\plotStartX + \startX - 0.5 * \specWidth - 0.25 * \specHeight, 1 * \specDist + \plotStartY + \startY +1 * \specHeight + 2 * \doaDist){\footnotesize frequency [kHz]};
% \node[anchor=center, rotate=90] at (\plotStartX - 0.86 * \specWidth, \plotStartY - 0.5 * \specHeight+ 1 * \specDist + 1*\specHeight) {\footnotesize Frequency in kHz};

% axis ticks
\foreach \x in {0, 1} {
    \foreach \y in {0, ..., 3}{
        \ifthenelse{\y = 0}{
            \node[anchor=north] at (\plotStartX + \startX +\x * \specWidth + \x * \specDist - \specDist, \plotStartY  - 0.5 * \specHeight - 2*\majorTick) {\footnotesize time [s]};
        }{}
        \foreach \t in {0, 2, 4, 6, 8, 10, 12} {
            \draw[line width=\tickWidth] (\plotStartX + \startX +\x * \specWidth + \x * \specDist + \t * \timeDist  - 0.5 * \specWidth, \startY + \y * \specDistY + \plotStartY + \y * \specHeight) --  (\plotStartX + \startX +\x * \specWidth + \x * \specDist + \t * \timeDist - 0.5 * \specWidth, \y * \specDistY + \plotStartY + \startY+ \y * \specHeight - \majorTick); 
            \ifthenelse{\y = 0}{ % this conditional throws an error, i dunno why
                \pgfmathsetmacro{\time}{int(\t * 0.5)} 
                \node[anchor=north] at (\plotStartX + \startX +\x * \specWidth + \x * \specDist + \t * \timeDist - 0.5 * \specWidth, \y * \specDistY + \plotStartY + \startY+ \y * \specHeight ) {\fontsize{\tickSize}{\tickSkip}\selectfont \time};
            }{
            }
            % doa label
            \ifthenelse{\t < 5}{
                \draw[line width=\tickWidth] (\plotStartX + \startX +\x * \specWidth + \x * \specDist  - 0.5 * \specWidth, \startY + \y * \specDistY + \plotStartY + \y * \specHeight + \t * \doaDist) --  (\plotStartX + \startX +\x * \specWidth + \x * \specDist - 0.5 * \specWidth - \majorTick, \y * \specDistY + \plotStartY + \startY+ \y * \specHeight + \t * \doaDist);
                \ifthenelse{\x = 0}{
                    \ifthenelse{\y = 3}{
                        % doa axis
                        \pgfmathsetmacro{\doa}{int(\t * 90 - 180)} 
                        \node[anchor=east] at (\plotStartX + \startX +\x * \specWidth + \x * \specDist - 0.5 * \specWidth, \y * \specDistY + \plotStartY + \startY + \y * \specHeight + \t * \doaDist){\fontsize{\tickSize}{\tickSkip}\selectfont \doa};
                    }
                    {
                        % freq axis
                        \pgfmathsetmacro{\freq}{int(\t * 2)}
                        \node[anchor=east] at (\plotStartX + \startX +\x * \specWidth + \x * \specDist - 0.5 * \specWidth, \y * \specDistY + \plotStartY + \startY+ \y * \specHeight + \t * \doaDist){\fontsize{\tickSize}{\tickSkip}\selectfont \freq};
                    }
                    {}
                }
                {}
            }
            {}
        }        
        \foreach \t in {1, 3, 5, 7, 9, 11} {
            \draw[line width=\tickWidth] (\plotStartX + \startX +\x * \specWidth + \x * \specDist + \t * \timeDist  - 0.5 * \specWidth, \startY + \y * \specDistY + \plotStartY + \y * \specHeight) --  (\plotStartX + \startX +\x * \specWidth + \x * \specDist + \t * \timeDist - 0.5 * \specWidth, \y * \specDistY + \plotStartY + \startY+ \y * \specHeight - \minorTick);
            \ifthenelse{\t < 5}{
                \draw[line width=\tickWidth] (\plotStartX + \startX +\x * \specWidth + \x * \specDist  - 0.5 * \specWidth, \startY + \y * \specDistY + \plotStartY + \y * \specHeight + \t * \doaDist) --  (\plotStartX + \startX +\x * \specWidth + \x * \specDist - 0.5 * \specWidth - \minorTick, \y * \specDistY + \plotStartY + \startY+ \y * \specHeight + \t * \doaDist);
            }
            {}
        }
    }
}

\end{tikzpicture}
\vspace*{-17.5pt}
\caption{Strongly (1,\,2) and weakly (6,\,7,\,8) guided extraction results with FT-\acs{jnf} \cite{tesch24ssf_journal} and varying tracking/training configurations during crossing of target ({\protect\tikz \protect\fill[tab_orange] (0mm,0mm) ++ (0mm,0.8mm) circle[radius=1pt];}) 
and interfering speaker ({\protect\tikz \protect\fill[tab_blue] (0mm,0mm) ++ (0mm,0.8mm) circle[radius=1pt];}).}
\label{fig:crossing_speaker}
\end{figure}

\textbf{\textit{Strongly guided speaker extraction}}
To investigate the impact of moving speakers on the extraction capabilities of a \ac{ssf}, we supply FT-\acs{jnf} with continuous ground truth \ac{doa} information.
When training in a stationary setting, FT-\acs{jnf} displays a sharp performance decline across all metrics with increasing speaker motion (1).
The same network but trained on a dynamic setup, in which we uniformly draw the expected azimuth shift $\mathbb{E}\hspace*{-2pt}\left\{ |\Delta \theta_t| \right\}$ from the evaluation interval, is only slightly affected by the speaker's motion and results in improved metrics for all non-stationary cases (2).
Since FT-\acs{jnf} optimized in a motionless environment (1) can fully rely on distinct and stationary spatial features, the network is bound to overfit on this specific setup achieving a very high spatial selectivity \cite{tesch24ssf_journal}.
However, when dropping this constraint, the \ac{ssf} is not able to cope with the increasingly challenging acoustic characteristics as well as incapable of handling overlapping speakers due to solely relying on the provided spatial cue.
On the other hand, as shown in \cref{fig:crossing_speaker}, training the identical network in a dynamic setting, FT-\acs{jnf} is able to resolve the ambiguity and attenuate the interfering speaker sufficiently.
This demonstrates that although being conditioned on a \textit{spatial} cue, \acp{ssf} are also capable of distinguishing and leveraging \textit{temporal-spectral} characteristics.

\begin{figure}[t!]
\begin{minipage}{0.48\columnwidth}
    \centering
    \begin{tikzpicture}
    \begin{axis}[
         every axis/.append style={
            font=\trackingFont, 
        },
            width=\trackingWidth,
            height=\trackingHeight,
            ylabel={\acs{acc}\,[\%]\,$\rightarrow$},
            xlabel={exp. \acs{doa} shift $\mathbb{E}\hspace*{-2pt}\left\{ |\Delta \theta_t| \right\}$ \big[{\scriptsize $\frac{^\circ}{5\mathrm{s}}$}\big]},
            xmin=0, xmax=360,
            ymin=0, ymax=100,
            xtick={0, 90, 180, 270, 360},
            xticklabels={0, 90, 180, 270, 360},
            ytick={0, 25, 50, 75, 100},
            yticklabels={0, 25, 50, 75, 100},
            xticklabel style={yshift=2pt}, 
            yticklabel style={xshift=2pt},
            ylabel near ticks, 
            ylabel style={yshift=0.5mm, xshift=0mm, rotate=0, anchor=center},
            xlabel style={yshift=-1.5mm, rotate=0, anchor=center},
            major tick length=\trackingMajorTick,
            minor tick length=\trackingMinorTick,
            minor y tick num=1,
            minor x tick num=2,
            ytick pos=left,
            xtick pos=bottom,
            tick align=outside,
            tick style={line width=\trackingTickWidth, color=black},
            grid=both,
            legend style={
                draw=none,
                at={(0.68, 0.28)},
                legend columns=1,
                cells={anchor=west},
                anchor=south,
                fill=none,
                row sep=-1mm,
                font=\scriptsize,
                % draw=black, % Draw a box around the legend
                % fill=white, % Optional: Fill the box with white color
            },
        ]

        % particle filter
        \addplot[
            color=tab_cyan, mark=diamond*,
            mark size=\trackingMarkSize,
            solid,
        ]
        coordinates {
            (0, 100.0)
            (30, 52.564360567747926)
            (60, 44.68905380250431)
            (90, 40.12638351549114)
            (120, 37.189066541212064)
            (150, 34.16066885554695)
            (180, 32.069683154847304)
            (210, 29.757431693251036)
            (240, 28.004539639007763)
            (270, 26.138354100388632)
            (300, 24.891888704114038)
            (330, 23.504740826845584)
            (360, 22.042437822435744)
        };
        \addlegendentry{\acs{das}-\acs{pf}\cite{ward03basic_particle_filter}};

        % neural tracking
        \addplot[
            color=tab_pink,
            mark=*,
            mark size=\trackingMarkSize,
            solid,
        ]
        coordinates {
            (0, 99.02333074292959)
            (30, 96.92863664440239)
            (60, 94.57701832345126)
            (90, 92.40911092163537)
            (120, 90.32257047942804)
            (150, 88.36368367573417)
            (180, 86.49619806182409)
            (210, 84.46324084549589)
            (240, 82.50535717501305)
            (270, 80.77681266417123)
            (300, 78.80009237746368)
            (330, 77.19022202053272)
            (360, 75.39154863631043)
        };
        \addlegendentry{Proposed};
        
    \end{axis}

\end{tikzpicture}
\end{minipage}\hfill
\begin{minipage}{0.48\columnwidth}
    \centering
    \begin{tikzpicture}
    \begin{axis}[
         every axis/.append style={
            font=\trackingFont, 
        },
            width=\trackingWidth,
            height=\trackingHeight,
            ylabel={\acs{ae}\,[°]\,$\leftarrow$},
            xlabel={exp. \acs{doa} shift $\mathbb{E}\hspace*{-2pt}\left\{ |\Delta \theta_t| \right\}$ \big[{\scriptsize $\frac{^\circ}{5\mathrm{s}}$}\big]},
            xmin=0, xmax=360,
            ymin=0.1, ymax=100,
            xtick={0, 90, 180, 270, 360},
            xticklabels={0, 90, 180, 270, 360},
            ytick={0.1, 1, 10, 100},
            xticklabel style={yshift=2pt}, 
            yticklabel style={xshift=2pt}, 
            yticklabels={0.1, 1, 10, 100},
            ylabel near ticks,
            ylabel style={yshift=0mm, xshift=0mm, rotate=0, anchor=center},
            xlabel style={yshift=-1.5mm, rotate=0, anchor=center},
            major tick length=\trackingMajorTick,
            minor tick length=\trackingMinorTick,
            % minor y tick num=1,
            ymode=log,
            minor x tick num=2,
            ytick pos=left,
            xtick pos=bottom,
            tick align=outside,
            tick style={line width=\trackingTickWidth, color=black},
            grid=both,
        ]

        % neural tracking
        \addplot[
            color=tab_pink,
            mark=*,
            mark size=\trackingMarkSize,
            solid,
        ]
        coordinates {
            (0, 0.69976806640625)
            (30, 0.8939666748046875)
            (60, 1.028656005859375)
            (90, 1.141693115234375)
            (120, 1.2457122802734375)
            (150, 1.3439788818359375)
            (180, 1.446197509765625)
            (210, 1.5518035888671875)
            (240, 1.6641845703125)
            (270, 1.7665863037109375)
            (300, 1.89093017578125)
            (330, 1.9994125366210938)
            (360, 2.12554931640625)
        };
        % \addlegendentry{Static train};

        % particle filter start 
        \addplot[
            color=tab_cyan, 
            mark=diamond*,
            mark size=\trackingMarkSize,
            dashed,
        ]
        coordinates {
            (0, 0.0000000001)
            (30, 4.6028289794921875)
        };
        % \addlegendentry{Predictor train};        

        % particle filter
        \addplot[
            color=tab_cyan, 
            mark=diamond*,
            mark size=\trackingMarkSize,
            solid,
        ]
        coordinates {
            (30, 4.6028289794921875)
            (60, 5.983367919921875)
            (90, 7.086090087890625)
            (120, 7.94586181640625)
            (150, 9.167266845703125)
            (180, 10.200469970703125)
            (210, 11.513679504394531)
            (240, 12.824493408203125)
            (270, 14.266372680664062)
            (300, 15.610824584960938)
            (330, 17.274032592773438)
            (360, 19.517608642578125)
        };
        % \addlegendentry{Predictor train};        

        % neural tracking mae error
        \addplot [name path=st_lower, fill=none, draw=none]
        coordinates{
            (0, 0.3264923095703125)
            (30, 0.41644287109375)
            (60, 0.4704742431640625)
            (90, 0.5181312561035156)
            (120, 0.5592041015625)
            (150, 0.598114013671875)
            (180, 0.6384735107421875)
            (210, 0.6806182861328125)
            (240, 0.7209739685058594)
            (270, 0.7599945068359375)
            (300, 0.80853271484375)
            (330, 0.8492889404296875)
            (360, 0.8956413269042969)
        };
        \addplot [name path=st_upper, fill=none, draw=none]
        coordinates{
            (0, 1.22698974609375)
            (30, 1.6307830810546875)
            (60, 1.94927978515625)
            (90, 2.2317161560058594)
            (120, 2.4976043701171875)
            (150, 2.7576065063476562)
            (180, 3.028961181640625)
            (210, 3.3193206787109375)
            (240, 3.6305885314941406)
            (270, 3.912384033203125)
            (300, 4.261688232421875)
            (330, 4.561012268066406)
            (360, 4.918552398681641)
        };
        \addplot[color=tab_pink, fill opacity=0.2] fill between[of=st_lower and st_upper];

         % particle filter mae error
        \addplot [name path=st_lower, fill=none, draw=none]
        coordinates{
            (30, 1.7423858642578125)
            (60, 2.3048858642578125)
            (90, 2.700824737548828)
            (120, 3.0066375732421875)
            (150, 3.352569580078125)
            (180, 3.6381988525390625)
            (210, 4.0072021484375)
            (240, 4.329517364501953)
            (270, 4.727394104003906)
            (300, 5.027626037597656)
            (330, 5.4089813232421875)
            (360, 5.875663757324219)
        };
        \addplot [name path=st_upper, fill=none, draw=none]
        coordinates{
            (30, 10.942306518554688)
            (60, 17.07470703125)
            (90, 23.885574340820312)
            (120, 30.321273803710938)
            (150, 38.29621124267578)
            (180, 43.97709655761719)
            (210, 50.10300064086914)
            (240, 55.01167297363281)
            (270, 60.22642517089844)
            (300, 64.36721801757812)
            (330, 68.70263671875)
            (360, 74.08256530761719)
        };
        \addplot[color=tab_cyan, fill opacity=0.2] fill between[of=st_lower and st_upper];

    \end{axis}

\end{tikzpicture}
\end{minipage}
\vspace*{-17.5pt}
\caption{Accuracy (\acs{acc}) and median angular error (\ac{ae}) of \acs{tst} algorithms. Shaded areas indicate 25\% and 75\% quartiles.}
\label{fig:tst_performance}
\vspace*{-10pt}
\end{figure}
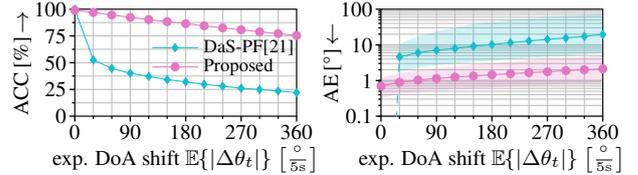

\textbf{\textit{Weakly guided speaker extraction}}
Instead of continuous oracle \ac{doa} information, our proposed \textit{weakly} guided \ac{tse} pipeline only relies on the target's initial \ac{doa} by utilizing an upstream \ac{tst} system.
To compare both tracking algorithms discussed in this work, we visualize sample trajectories in \cref{fig:moving_examples} and report frame-wise \ac{acc} with a 5° margin \cite{li23gcc-speaker, chen24locselect} and \ac{ae} in \cref{fig:tst_performance}.
Although the \ac{das}-\ac{pf} utilizes the motion model with oracle parametrization, which explains the perfect scores in the stationary case, the performance decreases rapidly with increasing speaker movement.
While our proposed network shows a much stronger and consistent overall performance, it also follows the same trend.
For this reason the simple concatenation of \ac{ssf} and \ac{tst} methods leads to a significant performance decrease in all cases (3,\,4,\,6,\,7).
However, with our proposed \textit{joint} training strategy, we are able to counteract this effect and even surpass the strongly guided \ac{tse} configuration in (1) regarding distortion and speech quality metrics for the non-stationary setups, underlining the importance of a dynamic dataset during training. 
\Cref{fig:crossing_speaker} illustrates the strong influence of erroneous \ac{doa} estimations from our \ac{tst} algorithm on FT-\acs{jnf} (6,\,7), which only the joint training (8) is able to resolve.
The biggest gain can be observed for the \ac{das}-\ac{pf} (5). 
Although providing highly inaccurate \ac{doa} estimates, the FT-\acs{jnf} is able to adapt with our joint training strategy and greatly improve across all metrics, see also \cref{fig:sisdr_performance}.

\section{Conclusion}
In this work we proposed a \textit{weakly} guided \ac{tse} pipeline which solely requires information about the target's initial direction.
By developing a synthetic dataset with continuous speaker movement, we could not only demonstrate the significant performance improvement of a commonly used \ac{ssf} in dynamic environments, but also its capability to resolve spatial ambiguities by learning to differentiate temporal-spectral patterns.
Finally, by introducing a \textit{joint} training strategy, we could surpass the performance of a mismatched, but strongly guided system with our own deep tracking \ac{nn} and also illustrate how \acp{ssf} can leverage highly inaccurate directional cues and achieve competitive performance even with a very simple tracking algorithm.

\section{Acknowledgments}
This work was supported by the Deutsche Forschungsgemeinschaft (DFG, German Research Foundation) under grant 508337379. Computational resources were provided by Regional Computer Center (RRZ) of the University of Hamburg (UHH) and Erlangen National High Performance Computing Center (NHR@FAU) of the Friedrich-Alexander-Universität Erlangen-Nürnberg (FAU) under the NHR project f104ac. NHR funding is provided by federal and Bavarian state authorities. NHR@FAU and RRZ hardware are partially funded by the DFG under grants 440719683 and 498394658 respectively.

\bibliographystyle{IEEEtran}
\bibliography{refs, strings}

\begin{thebibliography}{10}
\providecommand{\url}[1]{#1}
\csname url@samestyle\endcsname
\providecommand{\newblock}{\relax}
\providecommand{\bibinfo}[2]{#2}
\providecommand{\BIBentrySTDinterwordspacing}{\spaceskip=0pt\relax}
\providecommand{\BIBentryALTinterwordstretchfactor}{4}
\providecommand{\BIBentryALTinterwordspacing}{\spaceskip=\fontdimen2\font plus
\BIBentryALTinterwordstretchfactor\fontdimen3\font minus
  \fontdimen4\font\relax}
\providecommand{\BIBforeignlanguage}[2]{{%
\expandafter\ifx\csname l@#1\endcsname\relax
\typeout{** WARNING: IEEEtran.bst: No hyphenation pattern has been}%
\typeout{** loaded for the language `#1'. Using the pattern for}%
\typeout{** the default language instead.}%
\else
\language=\csname l@#1\endcsname
\fi
#2}}
\providecommand{\BIBdecl}{\relax}
\BIBdecl

\bibitem{cherry53cocktail_party}
E.~C. Cherry, ``Some experiments on the recognition of speech, with one and two
  ears,'' \emph{J. Acoust. Soc. Am.}, vol.~25, 1953.

\bibitem{zmolikova23tse_overview}
K.~Zmolikova, M.~Delcroix, T.~Ochiai, K.~Kinoshita, J.~Černocký, and D.~Yu,
  ``Neural target speech extraction: An overview,'' \emph{IEEE Signal Proc.
  Magazine}, vol.~40, 2023.

\bibitem{tesch24ssf_journal}
K.~Tesch and T.~Gerkmann, ``Multi-channel speech separation using spatially
  selective deep non-linear filters,'' \emph{IEEE/ACM TASLP}, vol.~32, 2024.

\bibitem{briegleb23icospa}
A.~Briegleb, M.~M. Halimeh, and W.~Kellermann, ``Exploiting spatial information
  with the informed complex-valued spatial autoencoder for target speaker
  extraction,'' in \emph{IEEE ICASSP}, 2023.

\bibitem{bohlender24sep_journal}
A.~Bohlender, A.~Spriet, W.~Tirry, and N.~Madhu, ``Spatially selective speaker
  separation using a {DNN} with a location dependent feature extraction,''
  \emph{IEEE/ACM TASLP}, vol.~32, 2024.

\bibitem{pandey12directional_speech_extraction}
A.~Pandey, S.~Lee, J.~Azcarreta, D.~Wong, and B.~Xu, ``All neural low-latency
  directional speech extraction,'' in \emph{Interspeech}, 2024.

\bibitem{gu24rezero}
R.~Gu and Y.~Luo, ``{ReZero}: Region-customizable sound extraction,''
  \emph{IEEE/ACM TASLP}, 2024.

\bibitem{vary06mvdr}
P.~Vary and R.~Martin, \emph{Digital Speech Transmission: Enhancement, Coding
  and Error Concealment}.\hskip 1em plus 0.5em minus 0.4em\relax Hoboken, NY,
  USA: Wiley, 2006.

\bibitem{chen20libricss}
Z.~Chen, T.~Yoshioka, L.~Lu, T.~Zhou, Z.~Meng, Y.~Luo, J.~Wu, X.~Xiao, and
  J.~Li, ``Continuous speech separation: Dataset and analysis,'' in \emph{IEEE
  ICASSP}, 2020.

\bibitem{barker18chime5}
J.~Barker, S.~Watanabe, E.~Vincent, and J.~Trmal, ``{The Fifth {'CHiME'} Speech
  Separation and Recognition Challenge: Dataset, Task and Baselines},'' in
  \emph{Interspeech}, 2018.

\bibitem{he243stse}
S.~He, J.~Liu, H.~Li, Y.~Yang, F.~Chen, and X.~Zhang, ``{3S-TSE}: Efficient
  three-stage target speaker extraction for real-time and low-resource
  applications,'' in \emph{IEEE ICASSP}, 2024.

\bibitem{bohlender23tracking_continuous_movement}
A.~Bohlender, L.~Roelens, and N.~Madhu, ``Improved deep speaker localization
  and tracking: Revised training paradigm and controlled latency,'' in
  \emph{IEEE ICASSP}, 2023.

\bibitem{meng22lspex}
M.~Ge, C.~Xu, L.~Wang, E.~S. Chng, J.~Dang, and H.~Li, ``{L-SpEx}: Localized
  target speaker extraction,'' in \emph{IEEE ICASSP}, 2022.

\bibitem{li19mc_speakerbeam}
G.~Li, S.~Liang, S.~Nie, W.~Liu, M.~Yu, L.~Chen, S.~Peng, and C.~Li,
  ``Direction-aware speaker beam for multi-channel speaker extraction.'' in
  \emph{Interspeech}, 2019.

\bibitem{li23gcc-speaker}
G.~Li, W.~Xue, W.~Liu, J.~Yi, and J.~Tao, ``{GCC-Speaker}: Target speaker
  localization with optimal speaker-dependent weighting in multi-speaker
  scenarios,'' in \emph{IEEE ICASSP}, 2023.

\bibitem{chen24locselect}
Y.~Chen, X.~Qian, Z.~Pan, K.~Chen, and H.~Li, ``{LocSelect}: Target speaker
  localization with an auditory selective hearing mechanism,'' in \emph{IEEE
  ICASSP}, 2024.

\bibitem{bohlender21ssl_temporal_context}
A.~Bohlender, A.~Spriet, W.~Tirry, and N.~Madhu, ``Exploiting temporal context
  in {CNN} based multisource {DoA} estimation,'' \emph{IEEE/ACM TASLP},
  vol.~29, 2021.

\bibitem{traa13wrapped_kalman_filter}
J.~Traa and P.~Smaragdis, ``A wrapped kalman filter for azimuthal speaker
  tracking,'' \emph{IEEE Signal Proc. Letters}, vol.~20, no.~12, 2013.

\bibitem{murase05kalman_moving_speaker_tracking}
M.~Murase, S.~Yamamoto, J.-M. Valin, K.~Nakadai, K.~Yamada, K.~Komatani,
  T.~Ogata, and H.~Okuno, ``Multiple moving speaker tracking by microphone
  array on mobile robot,'' in \emph{Interspeech}, 2005.

\bibitem{dong20pf_doa_coprime}
F.~Dong, L.~Xu, and X.~Li, ``Particle filter algorithm for {DoA} tracking using
  co-prime array,'' \emph{IEEE Comm. Letters}, vol.~24, 2020.

\bibitem{ward03basic_particle_filter}
D.~Ward, E.~Lehmann, and R.~Williamson, ``Particle filtering algorithms for
  tracking an acoustic source in a reverberant environment,'' \emph{IEEE Trans.
  on Speech and Audio Proc.}, vol.~11, 2003.

\bibitem{zhang17distributed_particle_filter}
Q.~Zhang, Z.~Chen, and F.~Yin, ``Speaker tracking based on distributed particle
  filter in distributed microphone networks,'' \emph{IEEE Trans. on Systems,
  Man, and Cybern.: Systems}, vol.~47, 2017.

\bibitem{revach22kalman_net}
G.~Revach, N.~Shlezinger, X.~Ni, A.~L. Escoriza, R.~J.~G. van Sloun, and Y.~C.
  Eldar, ``{KalmanNet}: Neural network aided kalman filtering for partially
  known dynamics,'' \emph{IEEE Trans. on Signal Proc.}, 2022.

\bibitem{chen21diff_pf_cond_norm_flow}
X.~Chen, H.~Wen, and Y.~Li, ``Differentiable particle filters through
  conditional normalizing flow,'' in \emph{IEEE FUSION}, 2021.

\bibitem{yang2022srp_dnn}
B.~Yang, H.~Liu, and X.~Li, ``{SRP-DNN}: Learning direct-path phase difference
  for multiple moving sound source localization,'' in \emph{IEEE ICASSP}, 2022.

\bibitem{yin22mimo_doanet}
H.~Yin, M.~Ge, Y.~Fu, G.~Zhang, L.~Wang, L.~Zhang, L.~Qiu, and J.~Dang,
  ``{MIMO-DoAnet}: Multi-channel input and multiple outputs {DoA} network with
  unknown number of sound sources,'' in \emph{Interspeech}, 2022.

\bibitem{xiong15gcc_phat}
X.~Xiao, S.~Zhao, X.~Zhong, D.~L. Jones, E.~S. Chng, and H.~Li, ``A
  learning-based approach to direction of arrival estimation in noisy and
  reverberant environments,'' in \emph{IEEE ICASSP}, 2015.

\bibitem{papageorgiou21ssl}
G.~K. Papageorgiou, M.~Sellathurai, and Y.~C. Eldar, ``Deep networks for {DoA}
  estimation in low {SNR},'' \emph{IEEE Trans. on Signal Proc.}, vol.~69, 2021.

\bibitem{panayotov15librispeech}
V.~Panayotov, G.~Chen, D.~Povey, and S.~Khudanpur, ``Librispeech: An {ASR}
  corpus based on public domain audio books,'' in \emph{IEEE ICASSP}, 2015.

\bibitem{allen79image_method}
J.~Allen and D.~Berkley, ``Image method for efficiently simulating small-room
  acoustics,'' \emph{J. Acoust. Soc. Am.}, vol.~65, 1979.

\bibitem{cosentino20librimix}
\BIBentryALTinterwordspacing
J.~Cosentino, M.~Pariente, S.~Cornell, A.~Deleforge, and E.~Vincent,
  ``Librimix: An open-source dataset for generalizable speech separation,''
  2020. [Online]. Available: \url{https://arxiv.org/abs/2005.11262}
\BIBentrySTDinterwordspacing

\bibitem{rong03survey_target_tracking}
X.~Rong~Li and V.~Jilkov, ``Survey of maneuvering target tracking. {Part I}.
  {Dynamic models},'' \emph{IEEE Trans. on Aerospace and Electronic Systems},
  vol.~39, 2003.

\bibitem{shimauchi14hann_window}
S.~Shimauchi and H.~Ohmuro, ``Accurate adaptive filtering in square-root {Hann}
  windowed short-time fourier transform domain,'' in \emph{IEEE ICASSP}, 2014.

\bibitem{wechsler24directional_filtering_directivity_control}
J.~Wechsler, S.~R. Chetupalli, M.~M. Halimeh, O.~Thiergart, and E.~A.~P.
  Habets, ``Neural directional filtering: Far-field directivity control with a
  small microphone array,'' in \emph{IWAENC}, 2024.

\bibitem{briegleb24constrained_vs_unconstained_filtering}
A.~Briegleb and W.~Kellermann, ``Spatially constrained vs. unconstrained
  filtering in neural spatiospectral filters for multichannel speech
  enhancement,'' in \emph{EUSIPCO}, 2024.

\bibitem{lentz24ftjnf_head_rotation}
B.~Lentz and R.~Martin, ``Utilizing head rotation data in {DNN}-based
  multi-channel speech enhancement for hearing aids,'' in \emph{IWAENC}, 2024.

\bibitem{diaz18gpu_rir}
D.~Diaz-Guerra, A.~Miguel, and J.~R. Beltr{\'a}n, ``{gpuRIR}: A {Python}
  library for room impulse response simulation with {GPU} acceleration,''
  \emph{Multimedia Tools and Applications}, vol.~80, 2018.

\bibitem{roux19sisdr_half_baked}
J.~L. Roux, S.~Wisdom, H.~Erdogan, and J.~R. Hershey, ``{SDR} – {Half-baked}
  or well done?'' in \emph{IEEE ICASSP}, 2019.

\bibitem{rix01pesq}
A.~Rix, J.~Beerends, M.~Hollier, and A.~Hekstra, ``Perceptual evaluation of
  speech quality {(PESQ)}-a new method for speech quality assessment of
  telephone networks and codecs,'' in \emph{IEEE ICASSP}, 2001.

\bibitem{jensen16estoi}
J.~Jensen and C.~H. Taal, ``An algorithm for predicting the intelligibility of
  speech masked by modulated noise maskers,'' \emph{IEEE/ACM TASLP}, vol.~24,
  2016.

\end{thebibliography}

\end{document}